\documentclass[sigconf]{acmart}

\def\BibTeX{{\rm B\kern-.05em{\sc i\kern-.025em b}\kern-.08emT\kern-.1667em\lower.7ex\hbox{E}\kern-.125emX}}
    
\acmYear{2019}
\acmConference[ICPP '19]{ICPP '19: International Conference on Parallel Processing}{August 5--8, 2019}{Kyoto, JP}

\usepackage{hyperref}
\usepackage[caption = false]{subfig}
\usepackage{microtype}
\usepackage{multirow}
\usepackage{graphicx}
\usepackage{epstopdf}
\usepackage[binary-units]{siunitx}
\usepackage{xcolor}
\usepackage{listings}

\definecolor{dkgreen}{rgb}{0,0.6,0}
\definecolor{gray}{rgb}{0.5,0.5,0.5}
\definecolor{mauve}{rgb}{0.58,0,0.82}

\begin{document}
%
\title[Optimized Performance of the
Hybrid MPI+MPI Parallel Codes]{MPI Collectives for Multi-core Clusters: Optimized Performance of the
Hybrid MPI+MPI Parallel Codes} 


%
\author{Huan Zhou}
\affiliation{%
  \institution{HLRS, University of Stuttgart}
  \streetaddress{Nobelstr. 19}
  \city{Stuttgart}
  \state{Germany}
  \postcode{70569}
}
\email{zhou@hlrs.de}

\author{Jose Gracia}
\affiliation{%
  \institution{HLRS, University of Stuttgart}
  \streetaddress{Nobelstr. 19}
  \city{Stuttgart}
  \state{Germany}
  \postcode{70569}
}
\email{gracia@hlrs.de}

\author{Ralf Schneider}
\affiliation{%
  \institution{HLRS, University of Stuttgart}
  \streetaddress{Nobelstr. 19}
  \city{Stuttgart}
  \state{Germany}
  \postcode{70569}
}
\email{schneider@hlrs.de}

\renewcommand{\shortauthors}{Zhou, Gracia and Schneider.}



%


\begin{abstract}

The advent of multi-/many-core processors in clusters advocates hybrid
parallel programming, which combines Message Passing Interface (MPI)
for inter-node parallelism with a shared memory model for on-node parallelism.
Compared to the traditional hybrid approach of MPI plus OpenMP,
a new, but promising hybrid approach of MPI plus MPI-3 shared-memory extensions
(MPI+MPI) is gaining attraction.
We describe an algorithmic approach for collective operations 
(with allgather and broadcast as concrete examples) in the context of hybrid MPI+MPI,
so as to minimize memory consumption and memory copies.
With this approach,
only one memory copy is maintained and shared by \mbox{on-node} processes.
This allows the removal of unnecessary on-node copies of replicated data
that are required between MPI processes when the collectives are invoked
in the context of pure MPI.
We compare our approach of collectives
for hybrid MPI+MPI and 
the traditional one for pure MPI,
and also have a discussion on the synchronization that is required to
guarantee data integrity.
The performance of our approach has been validated on a 
Cray XC40 system (Cray MPI) and NEC cluster (OpenMPI), 
showing that it achieves comparable or better performance 
for allgather operations.
We have further validated our approach with a standard computational
kernel, namely distributed matrix multiplication, and a Bayesian
Probabilistic Matrix Factorization code.
 
\end{abstract}


\maketitle

%

\section{Introduction}
MPI~\cite{MPISpec} has for decades retained its dominance in the circle
of the parallel programming model. It is widely used to write
portable and scalable parallel applications. Although MPI
was originally designed for distributed-memory systems where
\mbox{single-core} compute nodes were connected by a network,
it is also constantly tuned and adapted for gaining
acceptable performance on other types of systems, such as
on the shared memory symmetric multiprocessing (SMP)
and compound systems tightly coupled with SMP nodes.
Nowadays, the multi-core technology is extensively incorporated
into the current commodity supercomputers. Therefore,
the increased number of on-node data transfers
come along with the growing computational capability
of a single chip.
This will be certain to exacerbate memory bandwidth in the pure
\mbox{MPI-based} applications and then limit \mbox{per-core} affordable
problem sizes since the \mbox{intra-node}
data transfers are performed conventionally using the shared memory
as the transport layer.
As a result, the intention
of making full use of shared memory drives the programmers to
seek for the hybrid method which mixes the MPI
programming model (inter-node parallelism)
with a shared memory programming approach (on-node parallelism)
for allowing resources on a node to be shared.

Traditionally, the shared-memory models are OpenMP~\cite{OpenMP}
or Pthread~\cite{0201633922}.
This hybrid approach associates MPI rank with system-level or user-level
threads, which share memory inside the same address space.
OpenMP, as the most popular higher-level threading abstraction,
cooperates with MPI to benefit the parallel applications, especially
at high core counts where the scalability of pure MPI code suffers a lot
\cite{conf/pdp/RabenseifnerHJ09,conf/ipps/DrosinosK04}.
Importantly, a naive approach of migrating the pure MPI codes
to the hybrid MPI+OpenMP ones is to incrementally add OpenMP directives
to the computationally intense parts of the MPI application codes. 
This approach can produce serial sections that are only executed 
by the master thread as to an MPI process.
In this regard, such hybrid implementation may hardly outperform 
the pure MPI version, especially when the scaling of the MPI version is still good
\cite{techrep/hybridmpi+openmpi}.
Besides, the limitations of
the \mbox{thread-based} MPI libraries are fully
identified in a paper~\cite{conf/sc/FriedleyBHL13}.
Further, the \mbox{MPI-3} standard introduces a promising
process-based approach, namely the
MPI Shared Memory (SHM)
model~\cite{Introduction:SHM,conf/europar/ZhouIG15,conf/pvm/HoeflerDBBBBGKT12},
for supporting memory sharing between MPI processes within one node.
This leads to an innovative hybrid approach to parallel
programming with the combination of MPI and MPI SHM model~\cite{hoefler2013mpi+}.
The migration from the pure MPI codes to the hybrid MPI+MPI ones is 
straightforward since the work/data distribution across MPI processes 
(parallelism) stays unchanged. 
Even when the pure MPI applications are already good in scalability,
this hybrid scheme is expected to bring performance benefit due to 
the unchanged computationally parallelism 
and less communication overhead.
There are so far very limited experiences in writing efficient
hybrid MPI+MPI application codes, except for 
the study~\cite{hoefler2013mpi+} that presents
a hybrid MPI+MPI programming paradigm featuring point-to-point
communication operations (e.g., halo exchanges).
However, this paradigm is suboptimal since the halo zones are
needed for storing copies of replicated data for on-node processes.
Thus this does not comply with the hybrid programming scheme
requiring only one copy of replicated data between processes that are
eligible to share memory.

The MPI collectives (e.g., \mbox{\em MPI\_Bcast},
\mbox{\em MPI\_Allgather}, and \mbox{\em MPI\_Allreduce}) are important as they
are frequently
invoked in a spectrum of scientific applications or kernels~\cite{NASA}.
Their scalability is thus closely dependent on
the performance of MPI collective communication operations,
especially as the HPC systems continue to grow rapidly. Thus,
accelerating the collectives is a key to driving and achieving large-scale
scientific simulations.
Therefore, efficiently extending our experiences to include
the collective communication operations in the hybrid MPI+MPI context
is in high demand and appealing as well. 
Basically, the collectives in the pure MPI version give a copy
of the result to every on-node process, which should not be 
the case in the hybrid MPI+MPI version.

We (take allgather and broadcast as test cases) describe a 
hybrid implementation approach of collective operations in 
the hybrid MPI+MPI context, where only one copy
of replicated data is maintained within one node.
Meanwhile, we discuss the programmatic differences between 
the hybrid approach with the naive one in the pure MPI context,
under the condition that they perform the same work distribution
across cores (same parallelism in computation).
Unlike the naive approach, we need to take extra consideration
into the synchronization between on-node processes
for hybrid approach to guarantee the data integrity.
Further, we evaluate the time performance of the naive and 
hybrid approach (synchronization overhead inclusive)
for micro and application-level benchmarks.

The rest of the paper is organized as follows. 
In Section~\ref{sec:related-work} we give the related work and 
then introduce
the background knowledge about the shared memory model and 
the bridge communicator
in Section~\ref{sec:background}.
The detailed
implementation procedures of the hybrid collectives
in the MPI+MPI context are depicted in Section~\ref{sec:impl}.
The experimental results and analyses based on the micro and
application-level benchmarks are presented in Section~\ref{sec:eval}.
Section~\ref{sec:discussion} discusses two critical issues and 
Section~\ref{sec:conclusion} concludes.

\section{Related work}
\label{sec:related-work}

Traditionally collective algorithms have been well studied.
For each of collectives, many algorithms exist, where the different 
algorithms cater to different message sizes and numbers of processes\cite{thakur2005optimization}.
A lot of widely-used MPI implementations, such as MPICH~\cite{MPICH}, Intel MPI
\cite{IntelMPI} and OpenMPI~\cite{OpenMPI}, choose the most appropriate algorithm to use at runtime.

As the popularity of multi-core SMP clusters, much recent researches 
have been done in terms of SMP-aware (hierarchical) and shared memory collective
algorithm. SMP-aware algorithms are discussed to 
distinguish between intra-node and inter-node communications 
\cite{hasanov2015hierarchical, zhu2009hierarchical}.
In \cite{traff2014mpi},
the authors, take regular all-to-all for example, 
give an insight into the hierarchical design
of MPI collectives by solely using the MPI (3.0) functionalities
for the hierarchical communicator splitting.
The techniques for performing intra-node, shared-memory collectives
are investigated in \cite{conf/sc/FriedleyBHL13, conf/ppopp/FriedleyHBLM13, conf/hpdc/LiHS13}.
The hierarchical memory structure (multi-level caches) always
appears along with the multi-cores. 
 Several optimizations on the MPI collectives
are studied based on this characteristic to take advantage of
the shared cache as intra-node communication layer \cite{conf/pvm/GrahamS08, conf/ccgrid/MamidalaKDP08}. In \cite{mamidala2006efficient, qian2007rdma},
on top of SMP-aware, shared-memory is explored for intra-node communication
to improve performance and RDMA is used for the inter-node communication.
The above described SMP-aware collective algorithms are implemented based on
a single leader per node. A multi-leader-based allgather algorithm is then proposed 
to reduce the contention on memory and the leader process \cite{kandalla2009designing}.

\cite{conf/pvm/MaheoCPJ14}
optimized the MPI+OpenMP hybrid applications by using idle
OpenMP threads (refer to \cite{conf/ics/SiPBTI14}) to 
parallelize the invoked MPI collectives.

\section{Background}
\label{sec:background}
In this section, we briefly describe the MPI-3 shared-memory 
facilities and the two-level (intra-node and across-node)
of communicator splitting that is required 
for our work.

\textbf{MPI SHM model:}
\begin{figure}[tbp]
\begin{center}
\subfloat[Shared-memory communicator]{\label{fig:comm_split:sharedmem}\includegraphics[width=0.35\textwidth,height=0.14\textheight]{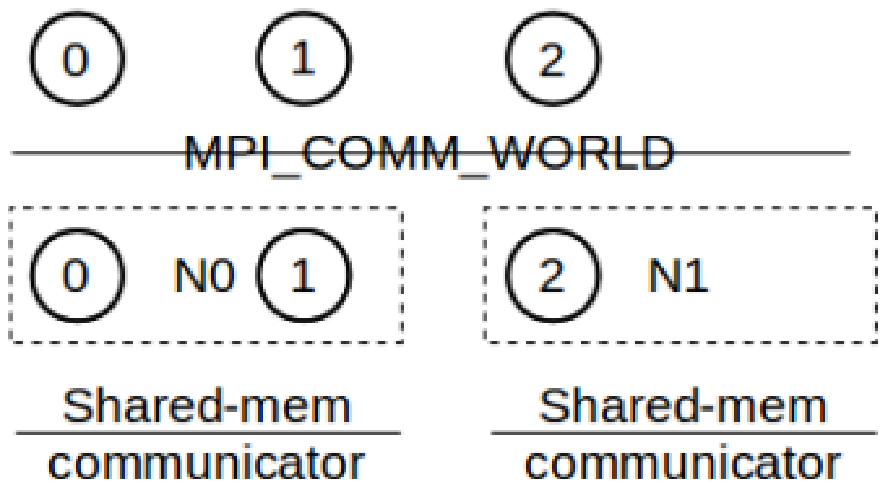}} \\
\subfloat[On-node memory sharing]{\label{fig:comm_split:onnode}\includegraphics[width=0.34\textwidth,height=0.12\textheight]{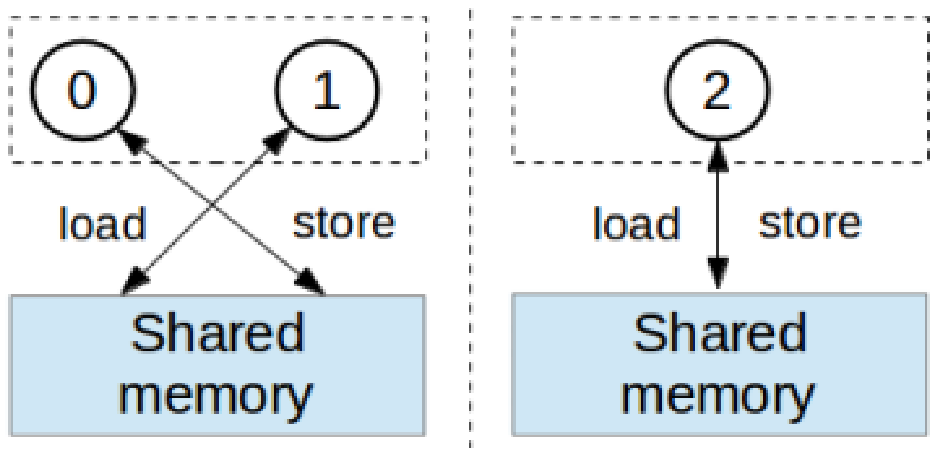}}

\end{center}
\caption{Shared-memory communicator splitting and shared-memory allocation.}
\label{fig:comm_split}
\end{figure}
The MPI-3 SHM facility plays a significant role in supporting the 
hybrid MPI+MPI programming model since it allows for the direct
load/store accesses to the shared data.
Figure~\ref{fig:comm_split:sharedmem} splits out two disjoint,
shared-memory sub-communicators  out of {\em MPI\_COMM\_WORLD} 
on two nodes 
by calling {\em MPI\_Comm\_split\_type} with 
the parameter of {\em MPI\_COMM\_TYPE\_SHARED}.
Each of the shared-memory communicators
identifies a group of  processes that can share data.
Shown in Figure~\ref{fig:comm_split:onnode},
a region of shared-memory is allocated on each of these
two shared-memory communicators via a call to \mbox{\em MPI\_Win\_allocate\_shared}.

Furthermore, the function {\em MPI$\_$Win$\_$shared$\_$query} is provided 
to query the base pointer to the memory on the target process. With the base pointer
the locally-allocated memory can be accessed by the MPI processes in the group of
the shared memory communicator with immediate
load/store operations (see Fig.~\ref{fig:comm_split:onnode}).

\textbf{Bridge communicator:}
\begin{figure}[tbp]
 \begin{center}
  \includegraphics[width=0.3\textwidth,height=0.14\textheight]{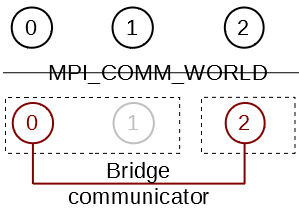}
 \end{center}
 \caption{The creation of bridge communicator}
 \label{fig:comm_split:node}
\end{figure}
Besides the above-mentioned shared-memory communicator, a bridge communicator
is needed for the across-node data exchanges in the hybrid MPI+MPI codes.
Shown in Figure~\ref{fig:comm_split:node},
one process on each node (the lowest ranking process)
is assigned as a \textit{leader} to take responsibility for
the data exchanges across nodes.
All the leaders constitute a bridge communicator, which is 
initially brought up for the hierarchical algorithms of MPI collectives
\cite{kandalla2009designing, traff2014mpi}. Also
\cite{traff2014mpi} presents a paradigmatic implementation of
splitting out the bridge communicator by using the facility provided by MPI-3
({\em MPI\_Comm\_split}).

\section{Implementation of MPI collectives in the hybrid MPI+MPI version}
\label{sec:impl}

In the pure MPI version, the collectives require a copy of the received data
to every process, 
which can, however, be omitted in the hybrid MPI+MPI version when
the process advances the following computation by merely using
the received data as an information without modifying it.
This is true in most of the
applications or kernels embracing the MPI collectives (refer to~\cite{NASA}).
Therefore, it is reasonable to only maintain one global copy of
the shared data, where the data to be sent is eligible to be 
shared by all on-node processes without the demand for intra-node 
memory copies. 
Not only is this implementation method 
expected to bring performance gain
but it also can 
save the \mbox{per-core} memory resources and 
keep the per-core memory costs to
constant as the number of the on-node processes increases.



Below we study the hybrid design of the widely-used collective
operations --
{\em MPI\_Allgather} and {\em MPI\_Bcast}
by assuming that the SMP-style rank placement is employed.
We then compare the implementation approaches of these two operations
between in the pure MPI and in the hybrid MPI+MPI version 
and in the end perform a detailed benchmarking study to assess
the benefits of using our approach over the naive one for
the pure MPI version.

\subsection{Allgather}
\label{sec:impl-allgather}
\begin{figure}[tbp]
\centering
 \subfloat[Allgather in the pure MPI version]{\label{fig:allgather:nativ}\includegraphics[width=0.4\textwidth,height=0.16\textheight]{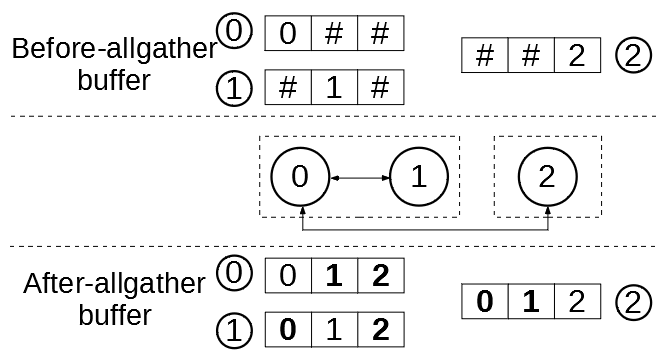}}\\
 \subfloat[Allgather in the hybrid MPI+MPI version]{\label{fig:allgather:hy}\includegraphics[width=0.35\textwidth,height=0.18\textheight]{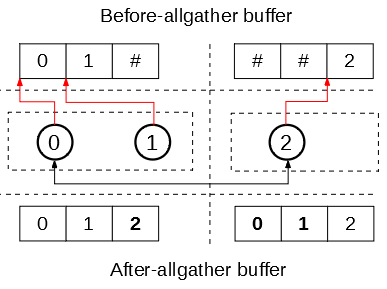}}
\caption{
Illustration on the allgather approaches in the context 
of pure MPI and hybrid MPI+MPI.
The notation $\sharp$ in the before-allgather buffer indicates an empty element.
The received elements in the after-allgather buffer are stressed with bold font.
The red arrow signifies a pointer while the black arrow denotes an inter-process communication.}
\label{fig:allgathernativehy}
\end{figure}

The all-to-all gather problem
is a typical MPI collective communication operation, where
each process in a given communicator distributes personal data
to all processes (including itself). 
Each process broadcasts data with the same size ({\em MPI\_Allgather}) in the
regular version, while the amount of data is not necessarily
equal for the irregular version ({\em MPI\_Allgatherv}~\cite{MPISpec}).

\begin{figure}[h!]
 \begin{lstlisting}[mathescape, escapechar=']
  /* Hierarchical communicator splitting '\cite{traff2014mpi}' */ 
  comm = MPI_COMM_WORLD;
  MPI_Comm_split_type(comm, 
                  MPI_COMM_TYPE_SHARED, 0, 
                  MPI_INFO_NULL, &sharedmemComm);
  MPI_Comm_rank(sharedmemComm, &sharedmemRank);
  leader = 0;
  MPI_Comm_split(comm, 
            (sharedmemRank==leader)?0:MPI_UNDEFINED,0,
            &bridgeComm);            
  MPI_Comm_size(comm, &nprocs);
  MPI_Comm_rank(comm, &rank);
  msgSize = (sharedmemRank==leader)?msg*nprocs:0;
  MPI_Win_allocate_shared(msgSize, sizeof(byte),
                      MPI_INFO_NULL, sharedmemComm,
                      &s_buf, &sharedmemWin);
  r_buf = s_buf;
  if (sharedmemRank != leader){
    MPI_Win_shared_query(sharedmemWin, leader, &s_buf);
  }
  s_buf = s_buf + msg*rank;
  s_buf[0..msg] = generate_random_character;
  if (bridgeComm != MPI_COMM_NULL){// Leaders
    if (bridgeCommSize > 1){// More than one node
      '\textcolor{red}{MPI\_Barrier(sharedmemComm)}';
      MPI_Allgatherv(s_buf,...,r_buf, bridgeComm);
      '\textcolor{red}{MPI\_Barrier(sharedmemComm)}';
    }
    else// Single node
      '\textcolor{red}{MPI\_Barrier(sharedmemComm)}';
  }
  else{// Children
    if (bridgeCommSize > 1){
      '\textcolor{red}{MPI\_Barrier(sharedmemComm)}';
      '\textcolor{red}{MPI\_Barrier(sharedmemComm)}';
    }
    else
      '\textcolor{red}{MPI\_Barrier(sharedmemComm)}';
  }
  Each process accesses the updated r_buf;
 \end{lstlisting}
\caption{Pseudo-code of the allgather implementation in the hybrid MPI+MPI version.}
\label{hybridallgather}
\end{figure} 

Figure~\ref{fig:allgathernativehy} illustrates the design difference of 
MPI collectives between in the pure MPI and hybrid MPI+MPI version.
Refer to  Sect.~\ref{sec:discussion} for the case
of the rank placement other than SMP-style.
The two legends in this figure 
are drawn to show rather the communication path and status
of on-node buffers than the switch of the allgather algorithms (recursive 
doubling or ring) in terms of different message sizes.
Fig.~\ref{fig:allgather:nativ} depicts the implementation algorithm
of allgather in the pure MPI version
and also marks the status of the buffer in
each process, which maintains its buffer with
the size of {\em MPI\_COMM\_WORLD} and works on it
independently. We suppose that the allgather shown
here is SMP-aware (cf. Sect.~\ref{sec:related-work}).
Initially, process i assigns a valid value to the i-th
element in its buffer as the local data, that is about to
be sent to other processes. During this allgather operation,
first the data is aggregated at the leader process,
and then the leaders exchange data, and finally, the leaders broadcast
data to their children.
After this operation, 
the content sent from each process is placed in rank order
in all processes' buffers and the copies of replicated data
are observed from the after-allgather buffers of processes 0 and 1.

The implementation of allgather in the
hybrid MPI+MPI version, shown in Fig.~\ref{fig:allgather:hy},
only demands one copy of buffer within one node,
which is allocated as a shared-memory segment.
The processes in the same node share this on-node buffer. Therefore,
only the leaders (process 0 and 2) of the two nodes exchange
all the valid elements when the \mbox{all-to-all} gather communication
operation is required.
In order to enable the
computational parallelism, a local pointer should be associated
with each process to indicate an element of the on-node shared buffer.
This element will be virtually viewed as the local data on which
the individual process can work independently.
In the example shown in Fig.~\ref{fig:allgather:hy}, process 0 and 1
point to the first and second element in the same buffer, respectively
and process 2 points to the third element in a different buffer.
Unlike the allgather in the pure MPI version, the aggregation and 
broadcast stages (intra-node message copies) are omitted in the 
hybrid MPI+MPI version. However, both of them face with the induced,
irregular problem over sets of irregularly populated 
node~\cite{traff2009relationships}. 
This is due to that the sent message size varies from
one node to another.
Intuitively the irregular allgather variant ({\em MPI\_Allgatherv})
is used for the across-node data transfers. This can also be replaced
by other regular operations (e.g., broadcast) for performance improvement.
In our approach, the irregular allgather variant is employed and evaluated
in Section below.

In Figure~\ref{hybridallgather} we list the 
allgather example written for the hybrid MPI+MPI version.
First, the hierarchical (shared-memory and bridge) 
communicator splitting~\cite{traff2014mpi} is performed from
line 2 to line 10. 
Then lines 11 through 20 allocate a contiguous region of
memory, that is shared among all processes in 
the given shared-memory communicator via a 
call to {\em MPI\_Win\_allocate\_shared}.
This function is invoked collectively with different memory sizes specified.
Each of the on-node leaders asks for
the contiguous memory that supposedly stores the
on-node shared data. On contrary, other on-node processes 
(viewed as \textit{children}) rely on
the invocation of \mbox{{\em MPI\_Win\_shared\_query}}
to query the base pointer
to the beginning address of the contiguous memory segment,
which is originally
allocated in the address space of the leader.
Note that the hierarchical communicator splitting and 
the allocation of the shared-memory segment are one-offs,
which can be amortized in the future by the repeatedly invoked 
{\em MPI\_Allgatherv} operation. 
Next, with this base pointer, each process can compute out the address
from which its local data starts and initializes its local
data independently in lines 21 and 22.
The local data of a process can also be viewed as its private data, on
which only this process is eligible to write. The private data
can also be shared (read) by other on-node processes.
Finally, data exchange across nodes (on the bridge communicator)
is performed from line 23 to line 39,
where the omitted computation of the parameters of the sets of
the received count and displacement for the irregular allgather
is also a one-off.
Besides the exchange, the on-node processes cooperate with each other by adding
synchronization calls -- barrier, which is specifically highlighted.
A barrier operation is added before and after the irregular
allgather operation, so as to guarantee data integrity.
The leaders wait on the first barrier until all the data partitions
are initialized by their children and 
the second barrier is invoked to make the children wait until
the leaders finish data exchanges. 
In the extreme case of one node, only one barrier
operation is performed on the shared-memory communicator to
assure that the one-node buffer is ready. 

  

\subsection{Broadcast}
\label{sec:impl-broadcast}
\begin{figure}[tbp]
\centering
 \includegraphics[width=0.38\textwidth,height=0.175\textheight]{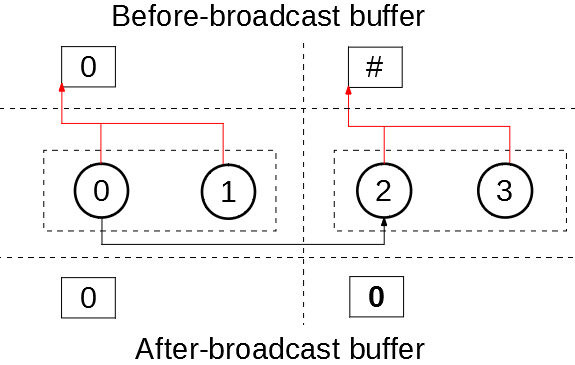}

\caption{
Illustration on the broadcast approach in the context
of hybrid MPI+MPI. The buffer above denotes before-broadcast buffer,
the buffer below denotes after-broadcast buffer.
Refer to Fig.~\ref{fig:allgathernativehy} for the 
the explanations of the notation $\sharp$, bold font
and arrows.}
\label{fig:allbroadcasthy}
\end{figure}

In a broadcast operation, a message is broadcast from the root
to all processes (itself included) of the given group.
Figure~\ref{fig:allbroadcasthy} shows the implementation algorithm
of the broadcast operation (with process 0 as the root)
in the hybrid MPI+MPI version. 
Unlike the pure MPI broadcast mechanism,
where each process allocates a memory buffer to store the broadcast data,
we maintain one shared memory segment
for the broadcast data within each node.
All processes within a node independently access the broadcast
data via a local pointer to the beginning
of this shared memory segment.
Here, performing the across-node broadcast operations over the process 0 and 2 is
straightforward since the size of the broadcast message remains the same as
the broadcast in the pure MPI version.

\begin{figure}[h!]
 \begin{lstlisting}[mathescape, escapechar=']
  if (rank == root)
    s_buf[0..msg] = generate_random_character;
  /* Broadcast across nodes */
  if (bridgeComm != MPI_COMM_NULL){// Leaders
    if (bridgeCommSize > 1){// More than one node
      MPI_Bcast(s_buf,..., root, bridgeComm);
      '\textcolor{red}{MPI\_Barrier(sharedmemComm)}';
    }
    else// Single node
      '\textcolor{red}{MPI\_Barrier(sharedmemComm)}';
  }
  else// Children
    '\textcolor{red}{MPI\_Barrier(sharedmemComm)}';
  Each process accesses the updated s_buf;
 \end{lstlisting}
\caption{Pseudo-code of the broadcast implementation in the hybrid MPI+MPI version.}
\label{hybridbcast}
\end{figure} 

Figure~\ref{hybridbcast} lists the 
pseudo code of the broadcast implementation in the hybrid MPI+MPI version. 
Fig.~\ref{hybridallgather} can be referenced for 
the omitted operations, such as hierarchical communicator splitting
and allocation of the shared-memory segment. 
However, unlike the allgather operation, 
only the root contributes to the broadcast data (see line 1-2).
Therefore,
the root other than other processes
is eligible to modify the shared data.
In this example,
one barrier operation is needed after the broadcast operation 
to guarantee that the broadcast data is ready
for all the on-node processes.

\section{Evaluation}
\label{sec:eval}

In this section, we compare the performance of the proposed approach of MPI
collectives for the hybrid MPI+MPI version with that of the naive
approach for the pure MPI version, based on micro and 
application-level experiments.
The micro experiments are conducted by measuring the allgather
latency for different schemes across different message sizes
and number of cores for different cluster configurations.
This micro-benchmark was modified from the OSU benchmark\footnote{http://mvapich.cse.ohio-state.edu/benchmarks/}
and averaged over 10000 executions. 

The two clusters are listed below:
\begin{enumerate}
\item Cray XC40 (aka. Hazel Hen):
Each node of Hazel Hen is a
\mbox{2-socket} system equipped with Intel Haswell
E5-2680v3 processors.
Each compute node has 24 cores running at 
\SI{2.5}{\giga\hertz} with
\SI{128}{\giga\byte} of DDR4 main memory.
The nodes are connected via the Cray Aries network which has a dragonfly topology.
Cray MPI library is employed on this cluster.
\item NEC cluster (aka. Vulcan): 
The architectures of node and processors are the same as Hazel Hen.
The nodes are connected via the InfiniBand network. 
OpenMPI library is employed on this cluster.
\end{enumerate}



The application-level experiments
consisting of SUMMA and BPMF, are conducted on the 
Cluster Hazel Hen (Cray MPI).
The overhead of the incurred synchronization calls is 
always included (explained in Sect.~\ref{sec:impl}) when
we evaluate the time performance of our hybrid approach
of MPI collectives. And
the extra one-off activities 
are not evaluated in this section.

\subsection{Allgather comparison}
\label{sec:eval-allgather}
\begin{figure}[tbp]
\begin{center}
\includegraphics[width=0.5\textwidth,height=0.21\textheight]{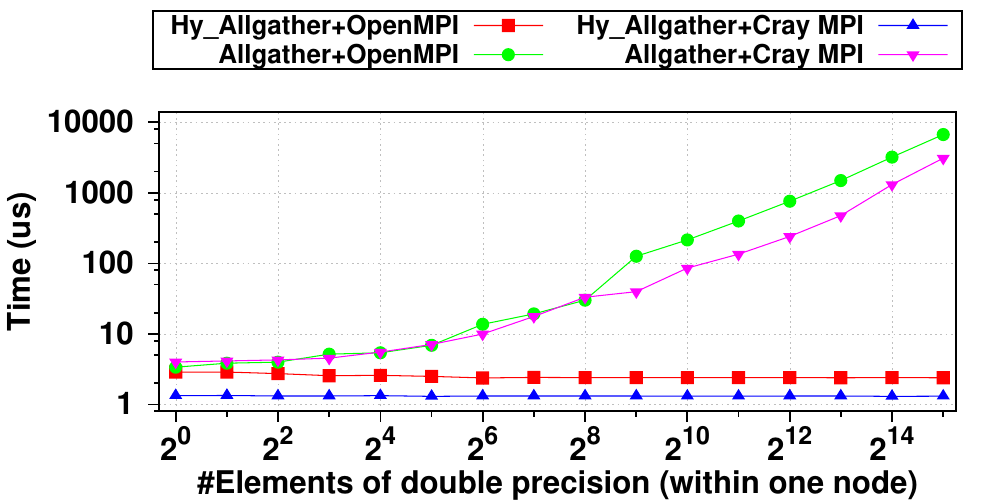}
\end{center}
\caption{Comparison results between \textit{Hy\_Allgather} and
\textit{Allgather} within one full node.
The number of the transferred elements is ranged from 1 to 32768.}
\label{allgather:onnode}
\end{figure}

\begin{figure*}[tbp]
\begin{center}
\subfloat[OpenMPI on Vulcan]{\label{allgather:1corepernode:OMPI}\includegraphics[width=0.5\textwidth,height=0.23\textheight]{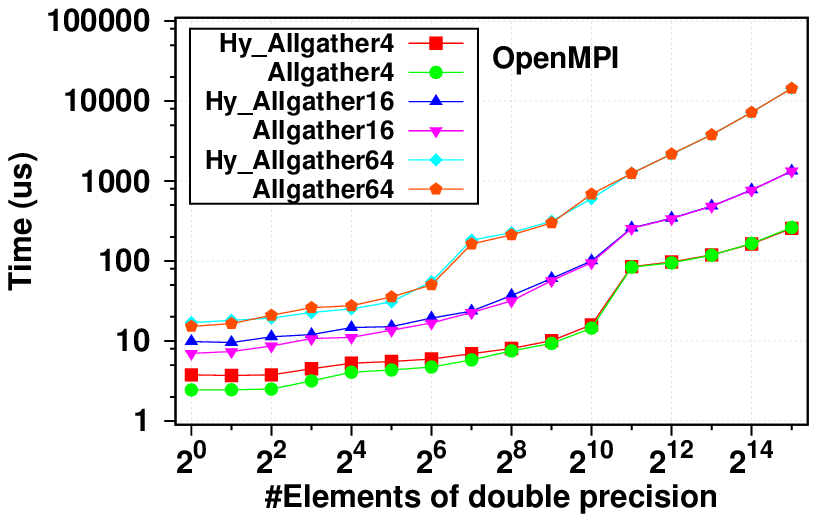}}
\subfloat[Cray MPI On Hazelhen]{\label{allgather:1corepernode:Cray}\includegraphics[width=0.5\textwidth,height=0.23\textheight]{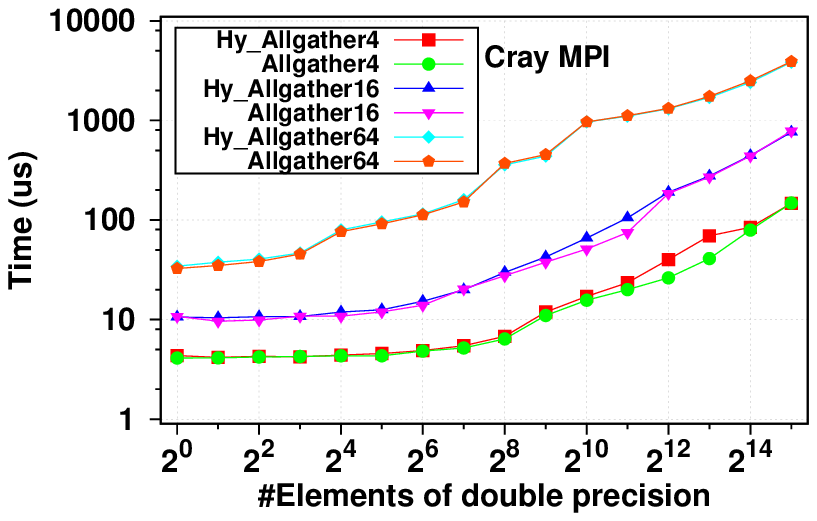}}
\end{center}
\caption{Comparison results between \textit{Hy\_Allgather}
and \textit{Allgather}
with 4, 16 and 64 nodes and one MPI process per node. 
The number of 
the transferred elements is ranged from 1 to 32768.}

\label{allgather:1corepernode}
\end{figure*}

\begin{figure*}[tbp]
\begin{center}
\subfloat[$\#$Elements of double precision 512]{\label{allgather:across64:512}\includegraphics[width=0.5\textwidth,height=0.21\textheight]{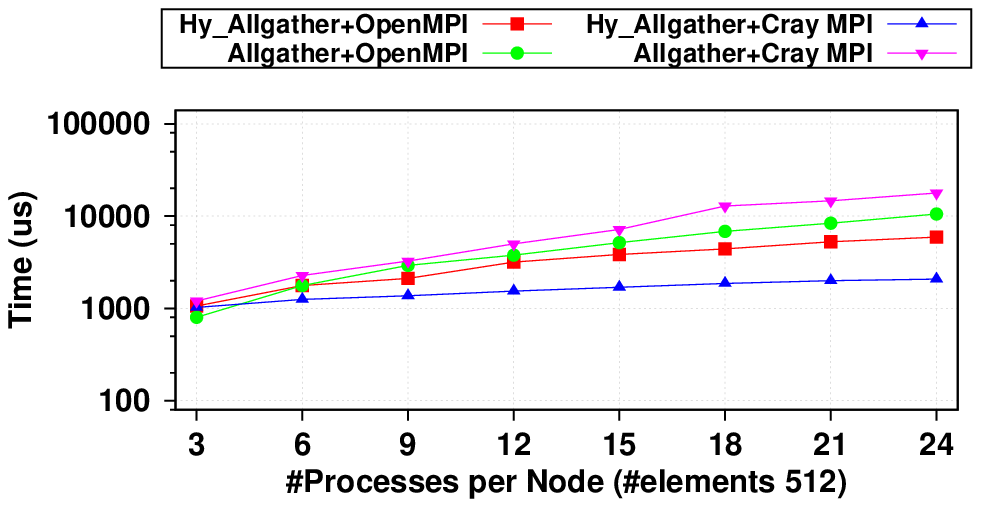}}
\subfloat[$\#$Elements of double precision 16384]{\label{allgather:across64:16384}\includegraphics[width=0.5\textwidth,height=0.21\textheight]{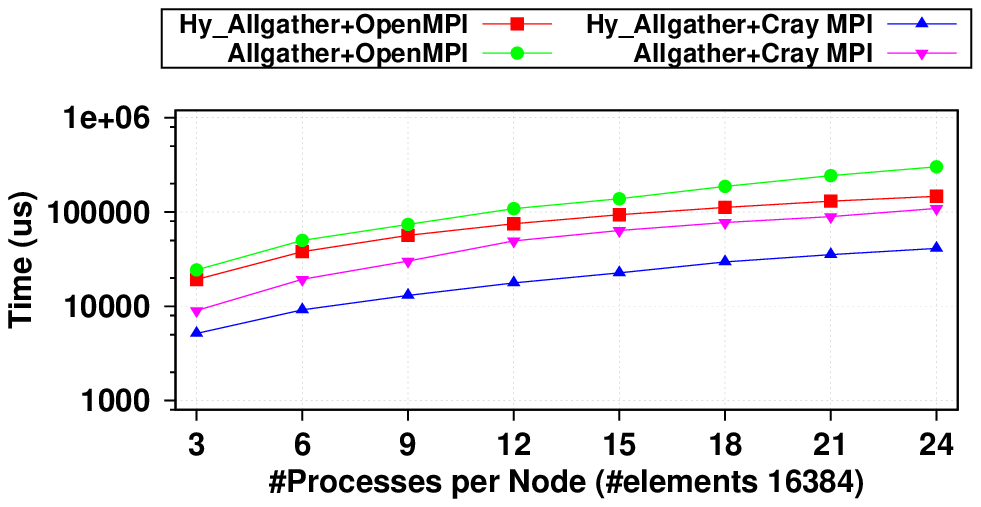}}
\end{center}
\caption{Comparison results between \textit{Hy\_Allgather}
and \textit{Allgather} across 64 nodes.
The number of MPI processes per node ranges from 3 to 24.}
\label{allgather:across64}
\end{figure*}

\begin{figure}[tbp]
\begin{center}
\includegraphics[width=0.5\textwidth,height=0.22\textheight]{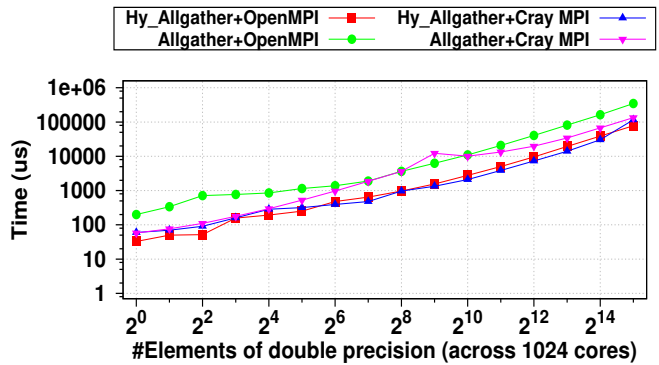}
\end{center}
\caption{Comparison results  between \mbox{\textit{Hy\_Allgather}}
and \textit{Allgather} on the irregularly populated nodes.}
\label{allgather:irregular}
\end{figure}

The abbreviations used for the allgather comparison
in these micro experiments are as follows:
\begin{itemize}
\item \textit{Hy\_Allgather}: Our Allgather approach for the hybrid MPI+MPI version, which
includes the synchronization calls.
\item \textit{Allgather}: The naive Allgather approach for the pure MPI version.
\end{itemize}

\subsubsection{Extreme cases}
\label{subsubsect:extreme}
We start with the investigation into the performance
comparison between our hybrid and naive allgather approach
under two extreme cases. In one scenario, we will have all MPI
processes residing on the same node. In the other scenario,
we allocate one MPI process per node, which means that
this comparison will equal to a comparison between 
{\em MPI\_Allgatherv} and {\em MPI\_Allgather}.
The count of the transferred elements of double precision floating point
(\SI{8}{\byte}) is increased from 1 to 32768.

In Figure~\ref{allgather:onnode}, we show the performance results
under the first extreme case, which is also the best case for our
approach since no inter-node data exchange is involved.
The curves for OpenMPI and Cray MPI are of the similar tendencies.
According to Fig.~\ref{hybridallgather}, 
it can be deduced that only an {\em MPI\_Barrier} is performed 
for our approach and as we expected, the overhead of \mbox{\textit{Hy\_Allgather}}
almost keeps constant.
The overhead of \textit{Allgather} goes up
steadily as the message size grows and is 
always greater than that of \textit{Hy\_Allgather}.


Figure~\ref{allgather:1corepernode} shows the performance results under second 
extreme case, which is also the worst case for our approach since it loses
the advantage offered by the on-node shared-memory mechanism.
Figure~\ref{allgather:1corepernode:OMPI}
and \ref{allgather:1corepernode:Cray} show the results running on the Vulcan with
OpenMPI library and Hazel Hen with Cray MPI library, respectively.  
This extreme experiment is conducted across 4, 16 and 64 nodes.
We can see that the time performance of \textit{Hy\_Allgather}
is slightly inferior to
that of \textit{Allgather} because the implementation of {\em MPI\_Allgatherv} 
is not as optimal as that of {\em MPI\_Allgather}\cite{traff2009relationships}. Further, it is obviously observed that the performance gap between 
\textit{Hy\_Allgather} and \textit{Allgather} shrinks
with 64 nodes. 

\subsubsection{General case}
In the plots shown in Sect.~\ref{subsubsect:extreme}, we can identify
the advantage of making use of on-node processes for our approach.
Besides, our approach suffers from the scheme of one MPI process
per node. Therefore, a further (more generalized) experiment is conducted to 
study how the advantage of \textit{Hy\_Allgather}
in performance
is affected by the number of processes per node.
Figure~\ref{allgather:across64} shows the results 
of this experiment with a fixed number of nodes of 64.
We can observe the growths of the performance advantage of \mbox{\textit{Hy\_Allgather}}
as the number of per-node-processes ranges from 3 to 24 for transferred element
count of 512 (see Figure~\ref{allgather:across64:512}) and 
16384 (see Figure~\ref{allgather:across64:16384}) respectively.
The \textit{Hy\_Allgather} starts to outperform \textit{Allgather} when the number 
of processes per node is beyond 3 for OpenMPI with 512 elements
and shows constant performance benefit in other situations. Moreover,
like the case of 512 elements, the \textit{Hy\_Allgather} for OpenMPI with
small message size (1 element) has outperformed
the \textit{Allgather} since the per-node-processes reached 3.


\subsubsection{Irregular case}

The experiments we discussed above cope with regular problems, where
the number of MPI processes is identical across all nodes.
However, \mbox{\em MPI\_Allgatherv} suffers performance penalty from the irregular problem since its performance is determined by the maximum amount of data to be received
by a node~\cite{traff2009relationships}.
Therefore,
Figure~\ref{allgather:irregular} compares the performance
between \textit{Hy\_Allgather} and \mbox{\textit{Allgather}} for an irregular
problem, which allocates 24 MPI processes on 42 nodes and 16
MPI processes on one node. We can discern the advantage 
of our approach due to its constant lower latencies.
In this section, only this irregular case is covered to
preliminarily demonstrate the performance superiority of our approach even
for the irregularly populated nodes.

\subsection{Application-level benchmarks}
\label{sec:eval-app}
In this section, we compare our approach of collectives
in the hybrid MPI+MPI version to that in the pure MPI version
for an application kernel of SUMMA and an application of BPMF.
The software environment used for the study
was CLE (Cray Linux Environment) with PrgEnv-gnu/5.2.82 and gcc version~6.2.0.

\subsubsection{SUMMA}
\label{sec:eval-SUMMA}
\begin{figure*}[htbp]
\begin{center}
\subfloat[8$\times$8]{\label{summa:8}\includegraphics[width=0.48\textwidth,height=0.24\textheight]{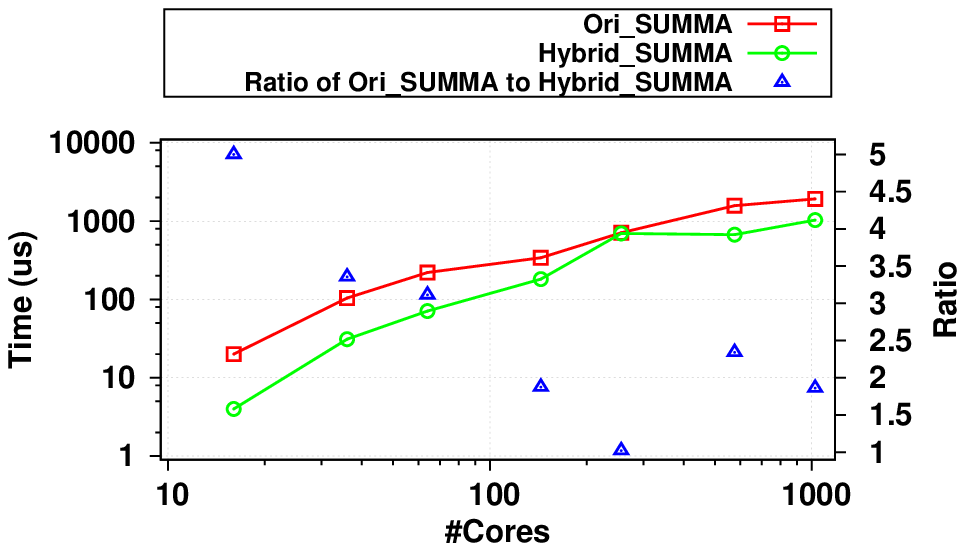}}
\subfloat[64$\times$64]{\label{summa:64}\includegraphics[width=0.48\textwidth,height=0.24\textheight]{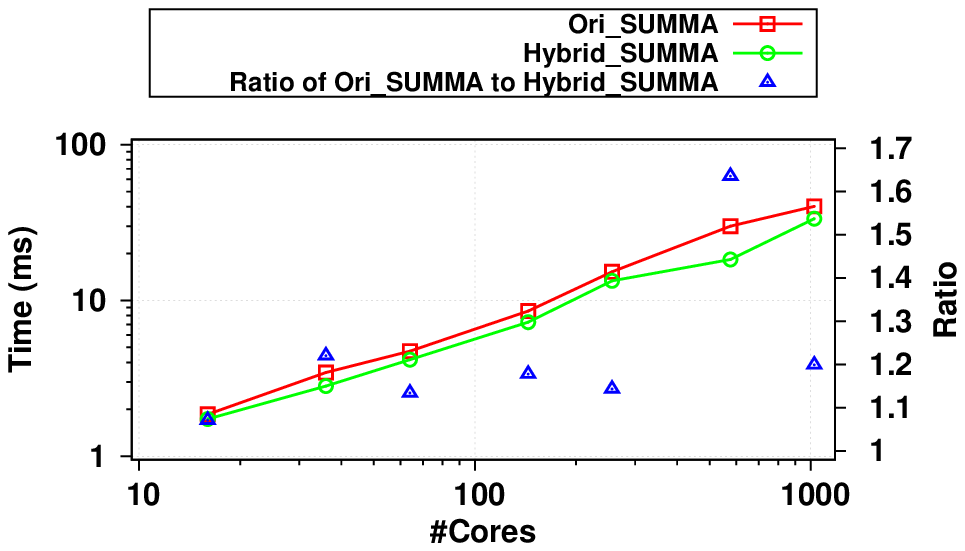}}\\
\subfloat[128$\times$128]{\label{summa:128}\includegraphics[width=0.48\textwidth,height=0.24\textheight]{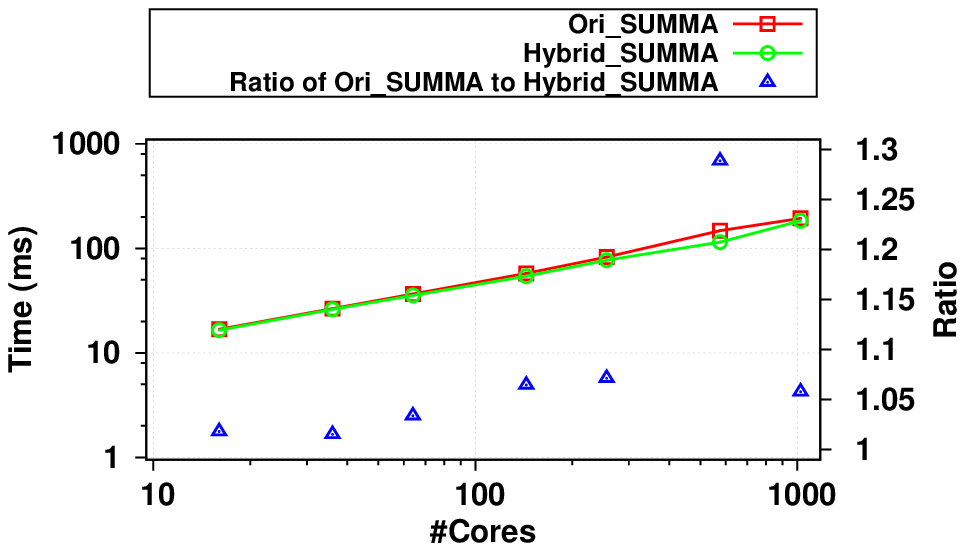}}
\subfloat[256$\times$256]{\label{summa:256}\includegraphics[width=0.48\textwidth,height=0.24\textheight]{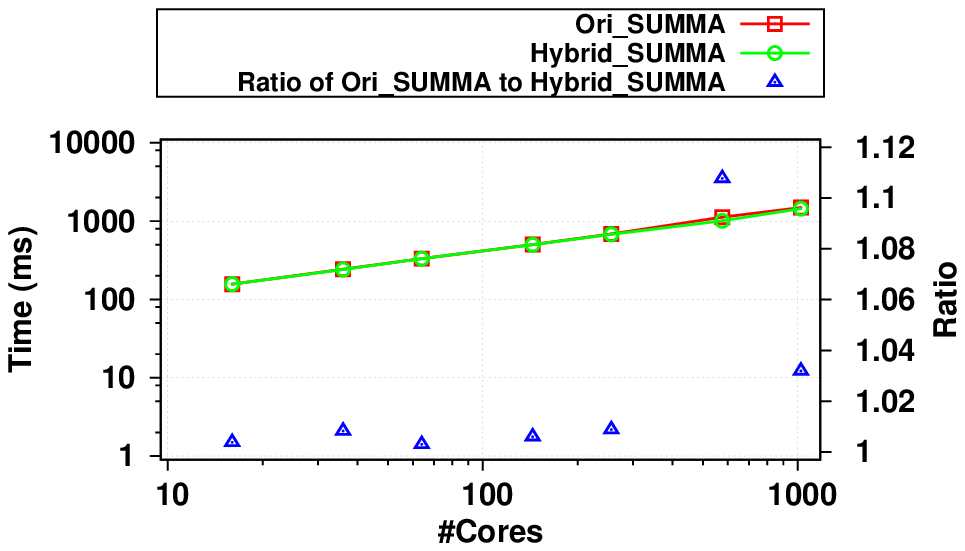}}\\
\end{center}

\caption{Performance comparison with respect to execution time (left y-axes) of the original SUMMA benchmark (red) and the version using hybrid collectives (green). Note, that the total problem size increases with number of cores and execution time is expected to scale as $\sqrt{\#cores}$. The ratio of execution times of the two versions (blue) is shown on the right y-axes. The panels correspond to different per-core matrix sizes as indicated.}    

\label{fig:summa}
\end{figure*}

SUMMA is a scalable universal algorithm for
implementing the dense matrix multiplication
\mbox{C= A $\times$ B}~\cite{journals/concurrency/GeijnW97}.
In this application kernel,
square matrices of size \mbox{N $\times$ N} are defined initially
for simplicity~\cite{journals/fgcs/Rico-GallegoML16}.
The elements of the matrices are then decomposed in blocks
of size \mbox{b $\times$ b} elements equally
on the $\sqrt{P}\times\sqrt{P}$ cores, where b equals to N/$\sqrt{P}$.
This application kernel consists of $\sqrt{P}$ iterations, where each iteration
triggers two broadcast communication operations of \mbox{b $\times$ b} data block
on the row and column communicators.
After obtaining the updated blocks, each process does its own computation.
This benchmark is averaged over at least 20 executions.

Figure~\ref{fig:summa} compares the performance between the 
SUMMA implementation using the naive broadcast 
approach in the pure MPI version ({\em Ori\_SUMMA})
and that using our broadcast approach 
in the hybrid MPI+MPI version ({\em Hy\_SUMMA}) with the per-core matrix sizes
of \mbox{8 $\times$ 8} (see Fig.~\ref{summa:8}),
\mbox{64 $\times$ 64} (see Fig.~\ref{summa:64}),
\mbox{128 $\times$ 128} (see Fig.~\ref{summa:128}) and
\mbox{256 $\times$ 256} (see Fig.~\ref{summa:256}).
The time performance ratios of
{\em Ori\_SUMMA} to {\em Hy\_SUMMA} are also provided in Fig.~\ref{fig:summa}.
Each matrix element is double precision float point
data type.
According to Fig.~\ref{hybridbcast}, a barrier synchronization across
the processes in the row or column communicator needs to be added 
after each of the two above mentioned broadcast operations in
{\em Hy\_SUMMA}.
The observed ratios that are consistently over one, demonstrate
the advantage of {\em Hy\_SUMMA} 
in time performance (compared to {\em Ori\_SUMMA})
with the increasing core count
for varying matrix sizes.
For small sizes, e.g., \mbox{8 $\times$ 8},
SUMMA is improved by as much as five times,
when all processes are arranged in the same node.
This is attributed to
the fact that the {\em Hy\_SUMMA} allows parallel computation
without any data movement in between.



\subsubsection{BPMF}
\label{sec:eval-BPMF}
\begin{figure}[tbp]
  \begin{center}
   \includegraphics[width=0.45\textwidth,height=0.22\textheight]{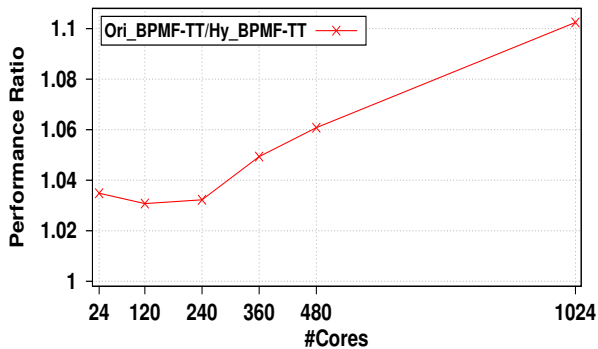}

\end{center}
\caption{Performance ratio of {\em Ori\_BPMF} to 
{\em Hy\_BPMF} as the number of the MPI processes (cores)
is increased from 24 to 1024,
for BPMF application.
\textit{TT} means the TotalTime
of the tested 20 iterations in BPMF application.}
\label{fig:comp_ratio}
\end{figure}

The BPMF (Bayesian Probabilistic Matrix
Factorization) application~\cite{ruslan:icml08, conf/cluster/AaCH16},
based on machine learning, can be used to predict
compound-on-target activity in chemogenomics.
The BPMF application works with the external library - Eigen.
Eigen is a high-level C++ library of template headers for linear algebra,
matrix and vector operations,
geometrical transformations, numerical solvers and related algorithms.
The BPMF version we used is an MPI code. We did 3 runs of the
application and averaged the execution times. The variance of timings
was small.
Within the application, the number of iterations to be sampled is set to be 20.
Each iteration consists of two distinct sampling
regions on movies and users.
Both regions end with the all-to-all gather communication routines.
Refer to the link~\footnote{https://github.com/ExaScience/bpmf/}
for more information about the BPMF code.

In this experiment we constantly use the chembl\_20 dataset as input data.
Below we measure the (strong scaling) performance of 
the BPMF application versions implemented with 
the naive approach for allgather ({\em Ori\_BPMF}) 
and our approach for it ({\em Hy\_BPMF}). 
Specifically, Figure~\ref{fig:comp_ratio} illustrates the
time performance ratio
of {\em Ori\_BPMF} to {\em Hy\_BPMF}.
According to 
Fig.~\ref{hybridallgather}, a barrier synchronization across
the on-node processes needs to be added before and after the 
all-to-all gather communication operations in {\em Hy\_BPMF}.
It is clearly observed that the ratios are always above 1,
which signifies that our allgather approach applied
in {\em Hy\_BPMF} can bring performance benefit for this BPMF application.
Besides, the ratio curve is on a slow
rise as the increasing number of cores.
It can be deduced that the profit from the adoption
of our allgather approach tends to grow steadily
as the process population increases.
Such incremental benefit can bring better scalability.
The ratio slightly increases (by 3.9\%)
with a total of 1024 cores.

Additionally, each of the two barrier calls works on
the processes in a shared-memory communicator (refer to Fig.\ref{fig:comm_split:sharedmem}), the size of which
will not exceed 24 (aka. the number of cores per node)
on Hazel Hen. Therefore, the extra overhead brought
by the synchronization is independent of the number of
MPI processes used, which further encourages the adoption
of our approach on large-scale computers.

\section{Discussion}
\label{sec:discussion}
In this section, two critical issues related to the synchronization
method and rank placement in our study are further addressed.

\textbf{Explicit synchronization:}
Synchronization and communication between processes are more decoupled in
hybrid MPI+MPI, compared to that in the pure MPI. In hybrid MPI+MPI,
synchronization operations are explicitly added to guarantee the 
data integrity and support a deterministic computation.
The synchronization mechanism comes in two flavors:

\begin{itemize}
\item Heavy-weight means: MPI Barrier across the on-node processes. This
is used in our paper and 
is also a common synchronization mechanism to collectively synchronize
among a set of processes.

\item Light-weight means: pairs of MPI point-to-point communications,
whereby one process can wait for another process to reach a given 
point in the program.
This is cheaper compared to Barrier and is basically 
used for the non-collective (regular or irregular) communication patterns.
\end{itemize}

\textbf{Rank placement:}
MPI processes are not necessarily mapped to the physical processors
in the SMP-style way, under which our approach for allgather is proposed.
In order to validate our approach for other rank placements than
SMP-style, the MPI derived datatype can be employed~\cite{traff2014mpi}.
However, the procedures of packing and unpacking always 
come with performance penalty.
Another way is to precompute the node-sorted global rank array~\cite{traff2014mpi},
with which we can deduce the corresponding
place of its block in the receive buffer 
in terms of any given global rank.

\section{Conclusion}
\label{sec:conclusion}

The prevailing application of multi-core technology encourages
economically memory usage and thus the hybrid approach combining
message passing (across-node) and shared memory (on-node).
MPI-3 Shared Memory (SHM) model emerges as an innovative
means of generating an efficient process-based hybrid 
MPI+MPI context on the multicore environment.
The MPI collective operation, especially \mbox{all-to-all}, however,
is not a scalable communication pattern. 
Therefore, adapting the implementation of 
MPI collectives to 
the hybrid MPI+MPI context is a key to obtaining 
a \mbox{memory-efficient} as well as
high performance scientific applications.
In this paper, we have proposed an
efficient approach implementing
the MPI collectives (such as {\em MPI\_Bcast}
and {\em MPI\_Allgather}) in the hybrid MPI+MPI context. 
This can fix the memory consumption issue at
the large-scale systems by only maintaining one copy
of replicated data shared by all processes within one node.
Besides, the occurrence of inter-process data transfers
is greatly reduced by removing the on-node memory copies.
Inserting synchronization calls appropriately is
necessary to guarantee the determinacy of the shared data.
Our approach of allgather was on par with or
outperformed that in the pure MPI context on Hazel Hen and Vulcan.
We did not discuss the time performance comparison 
of allgather for the message sizes larger than
\SI{256}{\kilo\byte},
where a pipeline method could, however, be applied~\cite{traff2008simple}.
Further, we observed consistent performance gains up to five times
for the SUMMA application kernel.
The total time of BPMF application using our allgather approach
fell by up to 10\%.

In our approach of MPI collectives for the hybrid MPI+MPI version,
the issue concerning the idle cores could also happen
when the leaders perform the inter-node data transfers
and their children wait on a barrier operation.
But this problem is lighter than
that in the MPI+OpenMP applications, where the OpenMP threads
are only active during the parallel computation phase and idle during
the MPI calls~\cite{conf/ics/SiPBTI14}.
Besides, from the programming perspective,
it is straightforward to let the on-node MPI processes
overlap with the network traffic by working on their own data regions.
Here, the synchronization
may be accelerated by using shared flags~\cite{conf/pvm/GrahamS08}
to signal other processes when required.

The hybrid MPI+MPI programming is promising but has not gained
the same attention as the hybrid MPI+OpenMP programming method,
since so far quite a few experiences on the writing of efficient hybrid 
MPI+MPI application codes are explored.
This paper can help to liberate the HPC programmers' mind from
the traditional multi-threaded MPI approaches when it comes to
multi-core environments, by highlighting an efficient approach for
implementing MPI collectives in the hybrid MPI+MPI context.
In addition, more experiences (e.g., p2p communications) are 
expected to popularize the implementation of 
the hybrid MPI+MPI application codes.

\section{Acknowledgments}

  The authors would like to thank Tom Vander Aa for offering
  the original code of BPMF application (aka. {\em Ori\_BPMF}).
  Part of this work was supported by European Community's Horizon 2020 POP project under    grant agreements n.~676553 and n.~824080.
  



 \bibliographystyle{ACM-Reference-Format}

   \bibliography{./paper}
%

%

\end{document}